# Plasmonic Beam Shaping in the Mid-infrared

**J. R. Pugh[1,2], J. Stokes[1], G.R.Nash[3], C. D. Stacey[2] and M. J. Cryan[1]**
[1] *University of Bristol, Bristol, BS8 1UB, UK, jon.pugh@bristol.ac.uk, m.cryan@bristol.ac.uk*
[2] *BAE Systems, Building 20R, Filton, Bristol, BS34 7QW, craig.stacey@baesystems.com*
[3] *University of Exeter, Exeter, EX4 4QJ, g.r.nash@exeter.ac.uk*

**Abstract:** Simulation results, fabrication details and measurements are presented for a one-dimensional aperture and grating array for the purpose of plasmonic beam shaping of the $\lambda=3.99\mu m$ output of an optically pumped semiconductor laser.
©2011 Optical Society of America

## 1. Introduction

Output beams of semiconductor lasers have the inherent disadvantage of a large divergence angle of around tens of degrees due to large diffraction caused by the small emission aperture of the devices [1]. The light output is mostly linearly polarized along a single direction, which is determined by the optical selection rules of the gain medium [2]. Laser beam shaping is conventionally conducted externally using optical components that can be bulky and expensive with operating wavelength limitations. In this study, Type-I [3] and Type-II [4] edge-emitting Optically Pumped Semiconductor Lasers (OPSLs), that operate based on interband transitions between the conduction and valence bands, are considered to demonstrate the role that plasmonics plays in providing a viable solution for beam shaping using corrugated thin metal films in the $3\mu m$ to $5\mu m$ wavelength range. Functional narrow-beam devices have been demonstrated further into the infrared (between $8\mu m$ and $10\mu m$) using Quantum Cascade Lasers (QCLs) by the Capasso group in Harvard using slit apertures in thin layers of gold surrounded by both 1D and 2D gratings [5-7]. Advantages of using the Type-I and Type-II OPSLs over the QCLs include higher operating efficiency at the shorter wavelengths required for gas detection and remote sensing of pollutants, free space optical communications and mid-infrared countermeasures. Cost of laser production is also significantly lower. This paper presents the design and fabrication of a slot and groove grating structure Focused Ion Beam (FIB) etched into a thin gold layer that has been deposited onto a $CaF_2$ glass slide. The output from the optically pumped Type-II QW edge-emitting device is reimaged onto the gold structure through the slide as a prototyping procedure allowing for small changes in grating design, which is now independent of the laser.

## 2. 2D Finite-Difference Time-Domain (FDTD) modeling and device design

The commercial FDTD modeling package Lumerical [8] has been used to optimize both the aperture and grating design for a plane wave gaussian modulated source centred at $\lambda=4\mu m$ placed inside the $CaF_2$ layer (n=1.41) and incident on the back surface of the gold layer as shown in figure 1(a). The gold layer contains a $1\mu m$ wide aperture surrounded on the output side by a 1D grating with groove width of 300nm and depth of 350nm. The distance from the centre of the aperture to the centre of the first grating slot, $d_1$, is $2.85\mu m$ and the remaining centre to centre groove spacing, $\Lambda$, is $3.95\mu m$ for a total of 64 periods each side of the aperture. The simulation space is divided into a 20nm square mesh and is surrounded by Perfectly Matched Layer (PML) boundaries. The monitor is a snapshot of the fields at $\lambda=3.99\mu m$ and stores all data necessary to compute a near-to-far field transform following the 3ps long simulation completion using a in-built function within Lumerical.

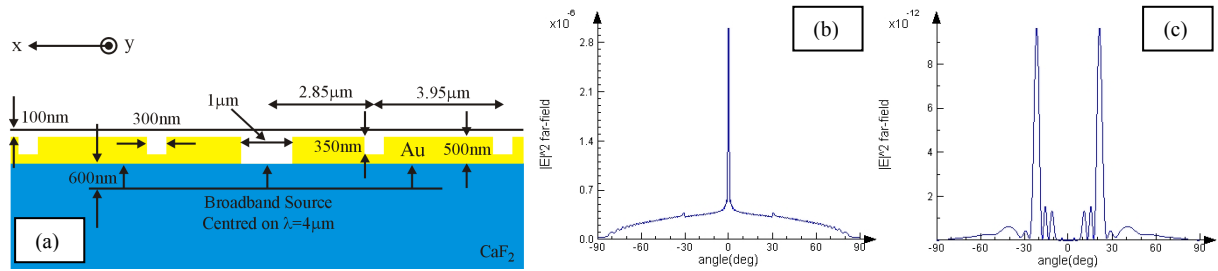

Fig. 1. (a) The aperture/grating design parameters. The $\lambda=4\mu m$ centred broadband plane-wave source is 600nm away from the metal input face and the monitor is a constant 100nm away from the metal output face. (b) shows the far field intensity profile with respect to angle where the source is polarized orthogonal to the aperture and grating (x-polarised), and (c) shows the case when it is parallel (y-polarised).

As the source is incident on the aperture, classical diffraction occurs along side the generation of surface plasmons (SPs) – collective oscillations of electrons in metal interacting with an electromagnetic field [9,10] – along the surface of the metal which contains grooves spaced by a distance approximately equal to the surface plasmon wavelength. The grating scattered light interferes with the diffracted light propagating through the aperture to form a very narrow beam, diverging by only an angle of ~1° in the far-field as shown in figure 1(b).

This is the case where the polarization of the source is in the growth plane, as is the case for type-I and type-II OPSLs the growth plane is the plane containing the diagram in figure 1(a)(x-z). The aperture and grating grooves are aligned so that their longer axis is orthogonal to the growth plane (y-direction). When the device is aligned so that the source polarization is parallel to the aperture and grating grooves (y-direction), the simulations yield intensity peaks in the far-field 300,000 times smaller than the orthogonal aligned polarization case as sown in figure 1(c).

## 2. Device fabrication and measurement

A schematic of the measurement setup is shown in fig. 2 along with a FIB image of the fabricated aperture and grating, which are both 50μm long. There are 10 periods of the grating on each side of the aperture. The pump laser – a Thulium Fiber laser with λ=1.94μm – is focused into a stripe using optics, including a cylindrical lens, onto the top surface of the Type-II OPSL, which is now situated inside a vacuum chamber and is cooled using a Stirling cooler down to ~60K. The output of the OPSL is measured to have an emission λ=3.99μm and fast-axis (y) divergence of 21.1º and slow-axis (x) divergence of 6.6º. The output profile of the OPSL is re-imaged onto the gold, through the $CaF_2$ slide using two 50mm focal length $CaF_2$ lenses. The aperture is aligned at the image focus and light propagating through the structure is collected using a further 30mm focal length lens. This light is then measured using a Spiricon Pyrocam III (SP-III) [11] which when coupled with a ZnSe window measures light incident on its detector array between λ=2-5μm. Fig. 3(right) shows the light measured by the SP-III for different orientations of the aperture and grating. The maximum intensity occurs when the aperture and groove long axis are aligned orthogonal to the OPSL polarization. As expected, as the slide is rotated so that its long axis is parallel with the OPSL polarization, very little light is measured. A larger amount is measured than is predicted in the modeling due to the finite length of the aperture.

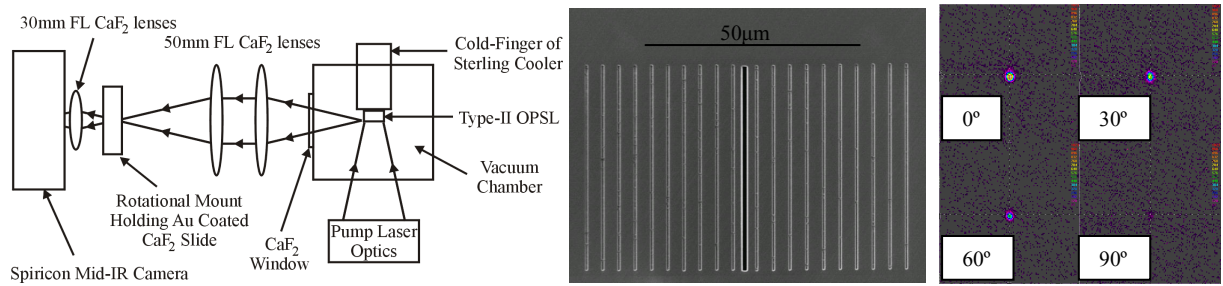

Fig. 2. (left) A schematic of the measurement setup, (middle) a FIB picture of the output side of the gold slide showing the cental aperture and grating, and (right) the light measured by the SP-III as the aperture and grating are rotated.

## 3. Conclusion

This paper presents the design and fabrication of a slot aperture surrounded by a 1D grating for the purpose of plasmonic beam shaping in the midinfrared. By optimizing the grating spacing around a 1μm wide aperture in a 500nm thick layer of gold, Lumerical FDTD simulations have shown that an incident plane wave at λ=4μm will give an output beam with a divergence of ~1º by a combination of diffraction and plasmonic effects. A simulation using gold gives a peak in intensity in the far field ~2.2 times larger than for a Perfect Electrical Conductor (PEC) where no plasmonic effects can occur. By rotating the aperture and grating relative to the source, it has been shown that maximum intensity occurs in the far-field when the source polarization is orthogonal to the grating axis and very little light is measured when it is parallel. Following further optimization of the Au coated $CaF_2$ prototyping procedure, Au will now be evaporated directly onto the OPSL facet and fully integrated "flat" lenses will be fabricated.